\documentclass[twocolumn,showpacs,preprintnumbers,amsmath,amssymb]{revtex4}

\usepackage{graphicx}

\begin{document}

\preprint{DF/IST-12.2003}

\title{
Electromagnetic radiation from collisions at almost the
speed of light: an extremely relativistic charged 
particle falling into a 
Schwarzschild black hole
}

\author{Vitor Cardoso}
\email{vcardoso@fisica.ist.utl.pt}
\affiliation{Centro Multidisciplinar de Astrof\'{\i}sica - CENTRA, 
Departamento de F\'{\i}sica, Instituto Superior T\'ecnico,
Av. Rovisco Pais 1, 1049-001 Lisboa, Portugal,} 

\author{Jos\'e P. S. Lemos}
\email{lemos@physics.columbia.edu} 
\affiliation{
Department of Physics, Columbia University, New York, NY 10027, 
USA, \&
Centro
Multidisciplinar de Astrof\'{\i}sica - CENTRA, Departamento de
F\'{\i}sica, Instituto Superior T\'ecnico, Av. Rovisco Pais 1,
1049-001 Lisboa, Portugal,} 

\author{Shijun Yoshida}
\email{yoshida@fisica.ist.utl.pt}
\affiliation{
{}Centro Multidisciplinar de Astrof\'{\i}sica - CENTRA, 
Departamento de F\'{\i}sica, Instituto Superior T\'ecnico,
Av. Rovisco Pais 1, 1049-001 Lisboa, Portugal.}

\date{\today}

\begin{abstract}
We investigate the electromagnetic radiation released during the high
energy collision of a charged point particle with a four-dimensional
Schwarzschild black hole.  We show that the spectra is flat, and well
described by a classical calculation.  We also compare the total
electromagnetic and gravitational energies emitted, and find that the
former is supressed in relation to the latter for very high energies.
These results could apply to the astrophysical world in the case
charged stars and small charged black holes are out there colliding
into large black holes, and to a very high energy collision experiment
in a four-dimensional world. In this latter scenario the calculation
is to be used for the moments just after the black hole formation,
when the collision of charged debris with the newly formed black hole
is certainly expected.  Since the calculation is four-dimensional, it
does not directly apply to Tev-scale gravity black holes, as these
inhabit a world of six to eleven dimensions, although our results
should qualitatively hold when extrapolated with some care to higher
dimensions.
\end{abstract}

\pacs{04.70.Bw, 04.30.Db}

\maketitle
\section{Introduction}

In a previous paper we have investigated the gravitational radiation
emitted by a relativistic particle infalling into a four-dimensional
Schwarzschild black hole \cite{cardosolemos0}, and afterwards extended
it to the infall in a Kerr metric spacetime \cite{cardosolemos1}.  The
interest of these studies lies in the fact they can describe the
gravitational radiation emitted in several collision phenomena, such
as the collision between small and massive black holes, between stars
and massive black holes, or between cosmic rays and small black holes,
to name a few.  The possibility of detecting gravitational radiation
by the several operating antennae is now real, and a closer
understanding of these collisional processes is accordingly important.
Of course, to fully understand collisional processes, one has in 
principle to go
beyond the infall of a particle and consider the collision between
two black holes with comparable masses, in which case one has to
take into account the full non-linearity and strong field regime of
Einstein's equations. Without resorting to numerical computations, 
but instead using topological arguments, Hawking was able 
to put  upper limits on the gravitational radiation emitted in
a black hole - black hole collsion in  four dimensions. This 
calculation was then refined by D'Eath and Payne \cite{payne}.
Their calculation made use of the fact that, if one boosts the
Schwarzschild metric to high velocities, then it approaches the
Aichelburg-Sexl \cite{aichelburg} metric, which is a shock wave
spacetime describing the gravitational field of a massless particle.
It was found that the total efficiency of the shock 
would be $\sim 16\%$ (see Yoshino and Nambu \cite{yoshino}
for the analysis in higher dimensions). In  \cite{payne}, 
the inclusion of a second term in the news function 
brings about a decrease in the efficiency from
$25\%$ (using only the first term) to $16\%$, 
so one is entitled to ask what degree of confidence
do these results and techniques offer.  For instance, 
one might suspect the third term to lower
even more the efficiency.  One needs to have other means 
of computing this process, since D'Eath and Payne's 
formalism is hard to pursue, due to its complexity. We have
recently \cite{cardosolemos0,cardosolemos1} proposed the ``point
particle paradigm'', which may well be a good candidate.  Handling
high energy collisions of point particles with black holes is rather
simple, since the point particle may be looked at as a perturbation in
the black hole spacetime, and the full machinery to handle black hole
perturbations is well developed \cite{bhperturbation}. The
interesting thing about the high energy collision of point particles
with black holes is that in the limit that the mass $\mu$ of the
particle goes to the mass $M$ of the black hole, the result for the
total radiated gravitational energy agrees well with other predictions
and never violates the area theorem
\cite{cardosolemos0,cardosolemos1}. Moreover, one can use this method
to treat the collision of rotating holes, which does not seem feasible
using previous techniques.

Up to now, most studies have concentrated on the emission of
gravitational radiation in the collision process, but if at least
one of the particles is charged, electromagnetic radiation is also
expected.  Astrophysically this is perhaps less relevant for the time
being, since the particles one can consider, such as stars or small
black holes, are in principle discharged, although it is possible the
latter ones can carry some charge (see, e.g., \cite{malheiro}). On the
other hand, from the point of view of elementary black holes such a 
study is certainly of interest: elementary black holes and
particles can easily carry electromagnetic charge, and the collision
between such objects can be put to operate, either by a machine in
some far future that can collide particles at center of mass energies
of the order of the four-dimensional Planck mass, $10^{16}\,$TeV, or
if there is any fundament in the TeV-scale gravity, by the near future
generation accelerators such as the LHC. The latter case has been
given a lot of consideration.  The TeV-scale gravity \cite{hamed}
requires extra large dimensions (of the order of sub-millimeters or
smaller) in order to lower the higher-dimensional Planck mass to
energies of the order of TeVs.  In these TeV-scale gravity models, the
gravitons are free to propagate in the higher dimensional spacetime,
whereas, due to experimental constraints, the standard model fields
live on a 3-brane, our universe.  In this scenario, a particle
collider with a center of mass energy of the order of TeVs can
copiously produce higher dimensional small spherical type black holes,
with radius of the order of fermis or less, in a space with large
extra dimensions \cite{bhprod}, or even black branes if some of the
extra dimensions are large and other small (see \cite{cav} for a 
review).  Moreover, in the collision process the gravitons
escape to the extra dimensions and are therefore much harder, harder than
usual, to detect. On the other hand, all the electromagnetic radiation
emitted in such a collision can be detected.

Thus it is important to know the electromagnetic spectrum and the
quantity of electromagnetic energy radiated in such an encounter.  In
this paper we extend the previous calculations into to the
electromagnetic window and find the electromagnetic radiation emitted
by a highly relativistic electrically charged particle infalling into
a four-dimensional Schwarzschild black hole.  This is a perturbation
calculation: we suppose a small charged particle falling into a
Schwarzschild black hole.  Some former works that have dealt with the
phenomenon of electromagnetic radiation from a charged particle
falling from infinity into a Schwarzschild black hole are
\cite{zerilli1,ruffini72,ruffini74,chandra}.  For example Ruffini
\cite{ruffini72} first calculated the electromagnetic energy spectra
and total radiated energy \cite{ruffini72}, for particles with low
Lorentz factors, $\gamma<<1$.  Here we go into high $\gamma$s.  We use
the point particle approximation and consider a charged point particle
colliding head-on at high velocity with a Schwarzschild black hole.
We show that the spectra is flat, and we also compare the total
electromagnetic and gravitational energies emitted, and find that the
former is supressed in relation to the latter for very high energies.
The numerical results extracted by us are in very good agreement with
Ruffini's results \cite{ruffini72}.  We shall also see that there is a
classical calculation for this process that agrees extremely well with
our numerical results.

Two comments are in order: 
(i) Our calculation is for the collision of a small particle with a
black hole in a four-dimensional world.  Therefore, in principle, it
could apply to the astrophysical world in the case charged stars and
charged black holes are out there, and to a very high energy collision
experiment in a four-dimensional world (without extra dimensions) and
with a fundamental Planck mass of $10^{16}\,$TeV. In this latter
scenario the calculation is to be used for the moments just after the
black hole formation, when the collision of charged debris with the
newly formed black hole is certainly expected.  Furthermore, the
calculation can be extended to a black hole - black hole collision as
we have already argued \cite{cardosolemos0,cardosolemos1}, and, in
addition, it can give clues, although it does not aplly directly to the 
usual collision process in a collider, i.e., the collision between one
non-charged particle with strong gravitational field (not necessarily
a black hole) and one charged particle. 
These results may, however, serve as
a model to the electromagnetic radiation emitted in the initial phase
of a newly formed black hole, the stage 
in which the black hole sheds its hair by emitting 
gauge radiation, such as electromagnetic radiation.
(ii) Our calculation is four-dimensional. It does not apply to
Tev-scale gravity black holes, since these inhabit a world of six to
eleven dimensions, with some of the dimensions being large, others
perhaps being small. However, although there 
are certainly some differences, qualitatively our results should 
hold when extrapolated with some care to higher dimensions.

\noindent
\section{Basic Formalism}
In this study, we assume the charged particle and the emitted radiation 
to be a small perturbation on the Schwarzschild spacetime, whose line 
element is given by
\begin{eqnarray}
ds^2=
-e^\nu\,dt^2+e^\lambda\,dr^2+r^2(d\theta^2+\sin^2\theta\,d\varphi^2) \,,
\label{metric}
\end{eqnarray}
where $e^\nu=e^{-\lambda}=1-2M/r$, and $M$ stands for the mass of the 
background spacetime. Because a complete prescription to deal with 
this problem was given by Zerilli \cite{zerilli1}, here we briefly show 
the essential parts of the formalism. 
In virtue of the static property and spherical symmetry of the background 
spacetime, perturbations can be decomposed by using vector harmonics 
for the angular variables $\theta$ and $\varphi$, and the Fourier components 
for the time variable $t$. After accomplishing a separation of variables, 
perturbations are therefore characterized by 
the harmonic indices $l$ and $m$, parity, and frequency $\omega$.
For the present case, where a particle 
falls straight into a black hole along the $z$-axis or the $\theta=0$ line, 
no axial parity perturbations are excited due to the symmetry of the motion. 
Thus, only the polar parity perturbations are considered in this study. 
According to Zerilli \cite{zerilli1}, our master equation for determining 
the electromagnetic radiation is given by a single wave equation,  
\begin{equation}
\frac{d^2 \tilde f_{lm}(\omega,r)}{ dr_*^2}+
\left[\omega^2-e^\nu\,\frac{l(l+1)}{ r^2}\right]\tilde f_{lm}(\omega,r)=
e^\nu \tilde S_{lm}\,,
\label{waveequation}
\end{equation}
where $r_*$ is the so-called tortoise coordinate, defined by
$dr/dr_*=e^{(\nu-\lambda)/2}$, or $r_*=r+2M\log(r/2M-1)$ in our case,
and $\tilde S_{lm}$ is the source term determined by the charge and
motion of the particle.  In the case of the radially falling particle,
$\tilde S_{lm}$ is generally given by
\begin{equation}
\tilde S_{lm}=-2q\sqrt{l+\frac{1}{2}}\,
\frac{e^{i\omega T(r)}}{ r^2}\,\delta_{m0}\,,
\end{equation}
where $\delta_{ij}$ means the Kronecker delta, and $q$ denotes the
charge of the particle. Here, the function $T(r)$ is the coordinate
time of the particle parameterized by the radial position $r$ of the
particle.  Since the charged particle is treated as a perturbation of
the background spacetime, the particle traces a radial geodesic of the
Schwarzschild geometry.  Thus, the function $T(r)$ is determined by
(see e.g., Chandrasekhar \cite{chandra})
\begin{equation}
\frac{dT(r)}{ dr}=-\frac{e^{(\lambda-\nu)/2}}{\sqrt{1-\gamma^{-2}\,e^\nu}}\,,
\end{equation}
where $\gamma$ is an integral of motion, given by
$\gamma=(1-v_\infty^2)^{-1/2}$, where $v_\infty$ is the radial
velocity of the particle at spatial infinity.  

Once solutions $\tilde f_{lm}(\omega,r)$ of equation (\ref{waveequation}) are 
calculated, time dependent functions $f_{lm}(t,r)$ can be obtained through 
the inverse Fourier transformation,  
\begin{equation}
f_{lm}(t,r)= \frac{1} {\sqrt{2\pi}}\int_{-\infty}^{\infty}e^{-i \omega t}
\tilde f_{lm}(\omega,r)d\omega\,.
\label{inversetransform}
\end{equation}
From the functions $f_{lm}(t,r)$, all the components of the field strength
$F_{\alpha\beta}$ can be derived except at the position of the charged
particle.  For example, the $r$--$\theta$ component $F_{r\theta}$ is
given by
\begin{equation}
F_{r\theta}(t,r,\theta,\varphi)=e^{-\nu}\sum_{l,m}f_{lm}(t,r)
\frac{\partial Y_{lm}}{ \partial\theta}(\theta,\varphi)\,.  
\end{equation}

In order to obtain a unique solution of equation (\ref{waveequation}),
two boundary conditions must be specified, and physically acceptable
conditions are purely outgoing wave at spatial infinity and purely
incoming wave at the horizon, which are due to equation
(\ref{inversetransform}) given by
\begin{equation}
\tilde f_{lm}(r)\rightarrow \left\{
\begin{array}{ll}
A_{lm}^{\rm in}(\omega)\,  e^{-i\omega r_*} & {\rm as} 
\quad r_*\rightarrow -\infty\,,\\
A_{lm}^{\rm out}(\omega)\, e^{i\omega r_*} & {\rm as} 
\quad r_*\rightarrow \infty \,.
\end{array}
\right.
\end{equation}
where $A_{lm}^{\rm out}(\omega)$ and $A_{lm}^{\rm in}(\omega)$ are not
dependent on $r$.  Since we are interested only in solutions at
spatial infinity, what we have to do is to obtain $A_{lm}^{\rm
out}(\omega)$'s as functions of $\omega$ for our purpose of this
study. Adapting a standard Green's function technique, we can write
$A_{lm}^{\rm out}$ as the integral, given by
\begin{equation}
A_{lm}^{\rm out}(\omega)=\frac{1}{ 2i\omega C_l(\omega)}
\int_{2M}^{\infty}f_{{\rm L},l} 
\tilde S_{lm}dr\,,
\end{equation}
where $f_{{\rm L},l}$ is a homogeneous solution of equation
(\ref{waveequation}) satisfying a boundary condition given by 
\begin{equation}
f_{{\rm L},l}\rightarrow \left\{
\begin{array}{ll}
e^{-i\omega r_*} & {\rm as} \quad r_*\rightarrow -\infty\,,\\
B_l(\omega)\, e^{i\omega r_*}+C_l(\omega)\, e^{-i\omega r_*} & 
{\rm as} \quad r_*\rightarrow \infty \,.
\end{array}
\right.
\label{boundary-conditions}
\end{equation}
The energy spectrum at spatial infinity is given by 
\begin{equation}
\frac{dE}{ d\omega}=\sum_l\frac{dE_l }{d\omega}=
\sum_l\frac{l(l+1)}{ 2\pi}\,|A_{l0}^{\rm out}(\omega)|^2 
\quad {\rm for} \quad \omega \ge 0\,, 
\label{energyspectrumnum}
\end{equation}
In order to obtain numerical values of $A_{lm}^{\rm out}(\omega)$, we
start the integration of $f_{L,l}$ and $\int_{2M}^rf_{{\rm
L},l}\tilde{S}_{lm}dr'$ at $r=2M(1+10^{-6})$ by using a Runge-Kutta
method. Those functions are then integrated out to large values of
$r$. The integration is stopped if the absolute value of
$\int_{2M}^rf_{{\rm L},l}\tilde{S}_{lm}dr'$ converges within the
required numerical accuracy, and we simultaneously match $f_{L,l}$
with an asymptotic solution satisfying condition
(\ref{boundary-conditions}) at spatial infinity, which is given in
\cite{chandra}, to obtain a value of $C_l(\omega)$.

\noindent
\section{Numerical Results}
\begin{figure}
\centerline{\includegraphics[width=8cm,height=8 cm]{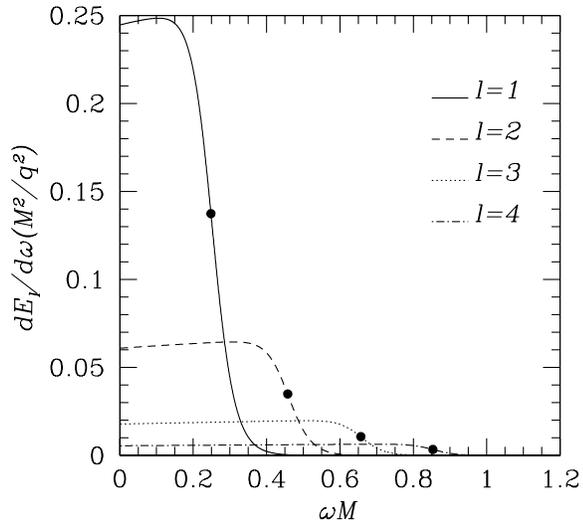}}
\caption{
The electromagnetic energy spectra for the four lowest radiatable
multipoles, for a $\gamma=2$ particle falling from
infinity into a Schwarzschild black hole. The filled circles on the
curves of the energy spectra indicate the frequency of the fundamental
quasinormal mode associated with the corresponding harmonic index $l$.
}
\label{spectragraph}
\end{figure}

\begin{figure}
\centerline{\includegraphics[width=8cm,height=8 cm]{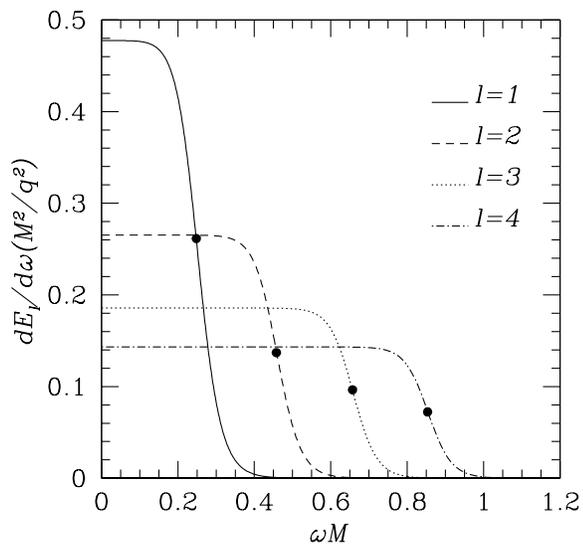}}
\caption{
The electromagnetic energy spectra for the four lowest radiatable
multipoles, for a $\gamma \gg 10$ particle falling from
infinity into a Schwarzschild black hole. The filled circles on the
curves of the energy spectra indicate the frequency of the fundamental
quasinormal mode associated with the corresponding harmonic index $l$.
}
\label{spectragraphinfty}
\end{figure}
Following the numerical procedure just outlined, we have computed the
energy spectrum for several values of $\gamma$.  In Figure 1 we show a
typical result of the energy spectrum, here for $\gamma=2$ and for the
four lowest radiatable multipoles. In Fig. 2 we show similar results
but for a particle with $\gamma \gg 10$.
A general feature of the energy
spectrum for high energy collisions is that it is flat up to some
critical frequency, after which it rapidly (exponentially) decreases
to zero.  This was also verified for the gravitational energy spectrum
resulting from the high-energy collision of a black hole with a point
particle \cite{cardosolemos0,cardosolemos1}. This critical frequency is given, 
in a very good approximation, by the frequency of the fundamental 
quasinormal mode 
associated with the corresponding harmonic index $l$. The corresponding 
quasinormal frequencies are also indicated by a filled circle in Figures 
1 and 2. 
The total energy $\Delta E$ radiated away is given by 
\begin{equation}
\Delta E=\sum_{l=1}^\infty\Delta E_l\,,
\label{total-energy}
\end{equation}
where $\Delta E_l$ is the total radiated energy associated with
$l$-pole electromagnetic radiation, which is given by the area under
the curve of the energy spectrum $dE_l/d\omega$ (see Figure 1).  For
the low $\gamma$ cases, the convergence of the infinite series
(\ref{total-energy}) is relatively good in the sense that a finite
summation of several $\Delta E_l$'s is sufficient to obtain a value of
$\Delta E$ with sufficient accuracy.  However, the convergence of the
total radiated energy gets worse with increasing $\gamma$.  This
contrasts with the analogous calculation of the gravitational energy
radiated in high energy collisions \cite{cardosolemos1} where a
rescaling was possible to show that $\Delta E \propto \gamma ^2$.
Here we shall see in the next section that the difficulty for the
electromagnetic case lies in the fact that $\Delta E \propto
\log{\gamma}$, for high $\gamma$'s.  For a very fast particle,
$\gamma\sim\infty$, it is quite difficult to numerically obtain the
total radiated energy, since all the $l$-pole energy spectra
$dE_l/d\omega$ make a substantial contribution to the total radiated
energy. In this study, we calculated the total energy for several
values of $\gamma$ with $\gamma\le 10$.  Some of these values are
shown in Table 1.

\begin{table}
\caption{\label{tab:zfl}  The total radiated electromagnetic
energy as function of $\gamma$ .}  
\begin{ruledtabular}
\begin{tabular}{c|c}
$\gamma$ & $\Delta E\,M q^{-2}$ \\ \hline
 1 &  $2.14\times 10^{-2}$   \\ 
 2 &  $1.14\times 10^{-1}$   \\ 
 3 &  $2.23\times 10^{-1}$   \\  
 4 &  $3.34\times 10^{-1}$   \\  
 5 &  $4.46\times 10^{-1}$    \\  
\end{tabular}
\end{ruledtabular}
\end{table}
%
Previous works concerned with the electromagnetic 
radiation from a charged particle falling from infinity into a Schwarzschild 
black hole include for example Ruffini \cite{ruffini72}.
Ruffini first calculated the 
electromagnetic energy spectra and total radiated energy \cite{ruffini72}, 
for low
$\gamma$ particles. 
The numerical results extracted by us are in very good agreement with Ruffini's
results \cite{ruffini72}.
\noindent
\section{The Classical Calculation}
There is a classical calculation in electromagnetism that suggests
itself as a model to use in the high energy collision of charged particles:
the radiation emitted when a charge, with constant velocity $v_\infty$ is 
suddenly decelerated to $v=0$. The deceleration is idealized as taking zero 
seconds. This model calculation has been applied with great success to, for
example, beta decay \cite{chang49,jackson}.
The result for the energy spectrum per solid angle is \cite{chang49,jackson} 
\begin{equation}
\left( \frac{d^2E}{ d\omega d\Omega}\right)_{\rm class}=
\frac{q^2v_\infty^2}{ 4\pi^2}
\frac{\sin^2\theta}{(1-v_\infty\cos\theta)^2}\,, 
\end{equation}
or, integrating over solid angle,  
\begin{equation}
\left(\frac{dE}{ d\omega}\right)_{\rm class}=
\frac{q^2}{ \pi}\left[\frac{1}{ v_\infty}
\log\left(\frac{1+v_\infty}{ 1-v_\infty}\right)-2\right]\,.
\label{dedw-flat}
\end{equation}
This formula has been tested with great accuracy experimentally. To get
the total energy, one has to integrate (\ref{dedw-flat}) over
frequencies, and a naive procedure would lead to infinities in the
total energy.  To obtain reasonable results, one has to impose a
cutoff $\omega_c$ in the frequency depending on the particular problem
under consideration, and one obtains
\begin{equation}
\Delta E_{\rm class}=\frac{q^2}{ \pi}\left[\frac{1}{ v_\infty}
\log\left(\frac{1+v_\infty}{ 1-v_\infty}\right)-2\right]\omega_c\,.
\label{totEclass}
\end{equation}
How well do these classical formulas fit into our numerical results?
Very well indeed. To allow a more direct comparison, we shall first
decompose the energy spectrum (\ref{dedw-flat}) into spherical
harmonics, i.e., into the energy spectrum associated with each
$l$-pole radiation as follows:
\begin{eqnarray}
\left(\frac{dE}{ d\omega}\right)_{\rm class}
=
\sum_l
\left(\frac{dE_l}{ d\omega}\right)_{\rm class} \nonumber \\
=
\sum_l\frac{q^2}{ 4\pi^2l(l+1)} 
\left|\int\frac{v_\infty\sin\theta}{ 1-v_\infty\cos\theta}
\frac{\partial Y_{l0}}{ \partial\theta}d\Omega \right|^2\,. 
\end{eqnarray}
When we consider the limit of $v_\infty\rightarrow 1$, as shown in
equation (\ref{dedw-flat}), the total energy spectrum $\left(\frac{dE}{
d\omega}\right)_{\rm class}$ diverges. However, the energy spectrum due to
$l$-pole radiation $\left(\frac{dE_l}{ d\omega}\right)_{\rm class}$ converges
to a finite value, given by
\begin{equation}
\left(\frac{dE_l}{ d\omega}\right)_{\rm class}= q^2\,\frac{2l+1}{ \pi l(l+1)}\,.
\label{fit0}
\end{equation}
This yields for example $\frac{dE}{d\omega}_{l=1}=0.477 q^2$ and
$\frac{dE}{d\omega}_{l=2}=0.265 q^2$ as $\gamma \rightarrow
\infty$. This may be compared with the numerical values in Fig. 2. The
numerical agreement is excellent not only for these lower multipoles,
but also for higher ones, and one has therefore established
numerically that for high energy collisions the classical result
(\ref{dedw-flat}) for the energy spectrum is valid.  Probably more
interesting is the total energy radiated. Since one already knows that
the classical result for $\frac{dE}{d\omega}$ is correct, let us now
try to predict the cutoff frequency by identifying $\Delta E$ in
equation (\ref{total-energy}) with (\ref{totEclass}).  Making use of this
definition of $\omega_c$ and numerical values of $\Delta E$, we
evaluate the cutoff frequencies $\omega_c$ in (\ref{totEclass}) 
for several values of $\gamma$, and
list the values of $\omega_c$ in Table II. In Figure 3, we also show
the cutoff frequencies as a function of $\gamma$. It is found in this
figure that for relatively high vales of $\gamma$, $\omega_c$ increase
almost linearly with $\gamma$, and a good fit to our numerical values
is given by 
\begin{equation}
M\omega_c=0.224+0.0598\,\gamma \,. 
\end{equation}
In Figure 3, we have shown this linear function, and we can confirm that 
this linear function is in good agreement with our numerical values for 
high $\gamma$.    
This means that a good approximation to the total electromagnetic energy
radiated is 
\begin{equation}
\Delta E= \frac{q^2}{M \pi}\left[\frac{1}{ v_\infty}
\log\left(\frac{1+v_\infty}{ 1-v_\infty}\right)-2\right](0.224+0.0598\gamma)\,, 
\label{eleapprox}
\end{equation}
or considering that $v_\infty \sim 1$ we have also 
\begin{equation}
\Delta E= \frac{2q^2}{M \pi}\left[\log{2\gamma}-1 \right]
(0.224+0.0598 \gamma)\,, 
\label{eleapprox2}
\end{equation}

\bigskip
\begin{figure}
\centerline{\includegraphics[width=8cm,height=8 cm]{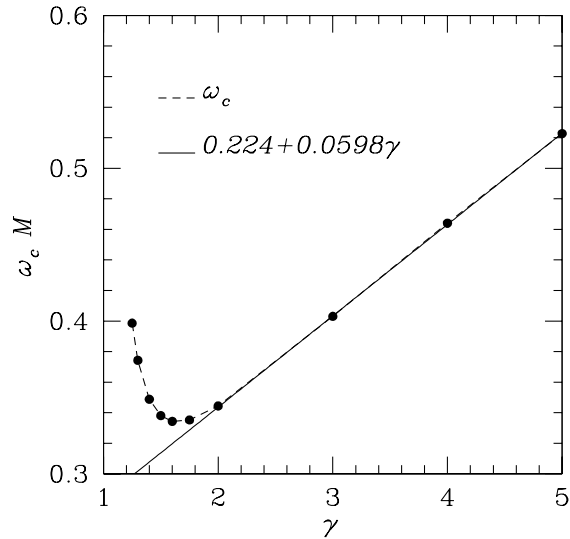}}
\caption{
The cutoff frequency, given as a function of $\gamma$. 
The filled circles on the curve of the cutoff frequency indicate the cutoff 
frequency calculated in this study. The linear function 
$\omega_c=0.224+0.0598\,\gamma$ is also shown. 
}
\label{cutofffrequencygraph}
\end{figure}
%
\begin{table}
\caption{\label{tab:zfl2}  The cutoff frequency $\omega _c$ as 
a function of $\gamma$ .}  
\begin{ruledtabular}
\begin{tabular}{c|c}
$\gamma$ &  $\omega _c\,M $ \\ \hline
 1 & $\infty$  \\ 
 2 & $3.44\times 10^{-1}$ \\ 
 3 & $4.03\times 10^{-1}$ \\  
 4 & $4.64\times 10^{-1}$ \\  
 5 & $5.23\times 10^{-1}$ \\  
\end{tabular}
\end{ruledtabular}
\end{table}

\noindent
\section{Comparing the Amount of Gravitational and Electromagnetic 
Energy radiated}
Previous studies on the high energy collision of point particles
with black holes led \cite{cardosolemos0,cardosolemos1} to the 
following conclusion:
if a high energy point particle of mass $\mu$ collides head on with
a mass $M$ non-rotating black hole, the total amount of gravitational
energy radiated is
\begin{equation}
\Delta E_g=0.26 \frac{\mu ^2 \gamma ^2}{M}\,,
\label{vlemos}
\end{equation}
where $\gamma$ is again the Lorentz factor for the point particle.
Furthermore, it was also found that the spectra is flat and
also well described by a classical calculation (here classical
means again working on a flat background). Now, for high $\gamma$'s 
(\ref{eleapprox2}) is 
\begin{equation}
\Delta E \sim \frac{0.038 q^2}{M }\gamma \log{2\gamma}\,,\,\,\gamma 
\rightarrow \infty\,. 
\label{eleapprox3}
\end{equation}
So, we get for the ratio electromagnetic energy over gravitational energy,
\begin{equation}
\frac{\Delta E}{\Delta E_g} \sim 0.146 (\frac {q}{\mu})^2 
\frac{\log{2\gamma}}{\gamma}\,,\,\,\gamma \rightarrow \infty\,. 
\label{ratio}
\end{equation}
This means that electromagnetic energy is supressed in relation
to gravitational energy, for very high energies.

\noindent
\section{Conclusions}
We have computed the electromagnetic spectrum and total energy
radiated during the high energy collision of a charged point particle
with a Schwarzschild black hole. Our results show that the
classical ``instantaneous collision'' calculation gives very good
results, in accordance with the full numerical ones.  We have dealt
only with zero impact (i.e., head-on) collisions, but these results
are easily generalized to non head-on collisions. For example, one
expects that the classical results \cite{chang49,jackson} hold for
such cases, and therefore that the total energy decreases as one
increases the impact parameter.  We stress that the
results presented here are valid for any particle charge $q$, 
as long as the total effective stress energy of the particle is
small compared to the total energy content of the black hole.
Previous works \cite{cardosolemos0,cardosolemos1} have shown that the
high energy collision between two black holes of equal mass may be
well studied through the collision of a point particle with a
black hole, and then taking the limit of equal mass, although this is
formally not allowed. It is our belief that the same may be done
here. One needs however some other method to attack this problem, 
to confirm or disprove this claim.  The investigation carried
here can also be carried over to higher dimensions, a case which is of
more direct interest to TeV-scale gravity scenarios. Since Standard
Model fields inhabit a four-dimensional brane, what one would need to
generalize this construction would be to modify the induced
four-dimensional metric (\ref{metric}), as was done for example in
\cite{kanti}. The generalization is straightforward.

\noindent {\bf Acknowledgements -}
We thank George Smoot for asking, during the talk of one of us (VC) in
the Fourth International Workshop on New Worlds in Astroparticle
Physics (University of Algarve, Faro, Portugal), how the gravitational
radiation emitted in the process compares with the electromagnetic
radiation, from which this work resulted.
This work was partially funded by Funda\c c\~ao para a Ci\^encia e
Tecnologia (FCT) -- Portugal through project CERN POCTI/FNU/49521/2002.
VC acknowledges finantial support from FCT through PRAXIS XXI
programme.
JPSL acknowledges finantial support from ICCTI/FCT and thanks
Observat\'orio Nacional do Rio de Janeiro for hospitality.
SY acknowledges finantial support from FCT through project SAPIENS
36280/99.



\end{document}